\documentclass[12pt]{article}
\usepackage{graphicx}
\begin{document}

\title{Market efficiency, anticipation and the  formation of bubbles-crashes}
\author{Serge Galam\thanks{serge.galam@polytechnique.edu},\\
Centre de Recherche en \'Epist\'emologie Appliqu\'ee,\\
\'Ecole Polytechnique and CNRS, \\CREA, Boulevard Victor, 32,
75015 Paris, France}
\maketitle

\begin{abstract}
A dynamical model is introduced for the  formation of a bullish or bearish trends driving an asset price in a given market. Initially, each agent decides to buy or sell according to its  personal opinion, which results from the combination of its own private information, the public information and its own analysis. It then adjusts such opinion through the market as it observes sequentially the behavior of a group of random selection of other agents. Its choice is then determined by a local majority rule including itself. Whenever the selected group is at a tie, i.e., it is undecided on what to do, the choice is determined by the local group belief with respect to the anticipated trend at that time. These local adjustments  create a dynamic that leads the market price formation.  In case of balanced anticipations the market is found to be efficient in being successful to make the ``right price" to emerge from the sequential aggregation of all the local individual informations which all together contain the fundamental value. However, when a leading optimistic belief prevails, the same efficient market mechanisms are found to produce a bullish dynamic even though most agents have bearish private informations. The market yields then a wider and wider discrepancy between the fundamental value and the market value, which in turn creates a speculative bubble. Nevertheless, there exists a limit in the growing of the bubble where private opinions take over again and at once  invert the trend, originating a sudden bearish trend. Moreover, in the case of a drastic shift  in the collective expectations, a huge drop in price levels may also occur extremely fast and puts the market out of control, it is a market crash.
\end{abstract}

Key words: renormalization group, sociophysics, opinion dynamic, finance

\newpage

\section{From the idea of each agent to the opinion of everyone}

In this paper, I apply sociophysics \cite{santo, eco, review} to the analysis of the aggregation of opinions in financial markets. A simple model is shown to describe the movement of a stock price far away from its fundamental value, through a process of iterations of individual decisions made by agents which interact by groups of odd or even size. Under certain social psychological configurations of the market, and depending on the parity of the size of the local groups, individual opinions combination can lead the stock price to reflect its fundamental value or be fundamentally different from it \cite{walter}.

A theoretical basis  is thus provided to Keynes view that stock prices can and actually differ from their fundamental values \cite{key1,key2}. Some models used in physics appear to be well adapted to the formalization of this perspective \cite{1,2,3,4} along a huge collection of works in economy \cite{5,6,7,8} with much attention being devoted to the formation of bubbles and crashes \cite{9,10,11,12,13,14,15}.

The novelty of the model  presented here  lies in the use of a single frame which embodies both the market efficiency and the formation of bubble as well as their burst. Although very simple, my scheme is able to explain two important insights by Keynes on how individual opinions combine and lead to a collective bullish or bearish behavior, and how expectations by agents can, in some cases, compromise the efficiency of the market pricing process, then reducing drastically the impact of individual information on the global outcome of the market process. However the distortion of the market efficiency is bounded and beyond some gap, the efficiency is brutally restored spontaneously. The model relies one my earlier works on opinion dynamics in social issues \cite{mino, hetero}.

 \subsection{Microstructure of the market and aggregation of individual opinions using the local sharing of informations}

Generally speaking, this approach is part of a broader increased attention to the aggregation of individual opinions in the last twenty years or so, especially on market microstructure and the way personal individual information are combined by the market equilibrium. Generally, these works are based on the hypothesis of lags in knowledge acquisition between agents. This leads them to influence and imitate each other in featuring mimetic behaviors. In this context, sociophysics  adds to these works by analyzing decision-making processes among groups of individuals with financial operators being divided into local groups like traders, portfolio managers, arbitrageurs and others.  The model explains how the structure of these groups governs the way of information processing by the market.

As done by usual models of market microstructure \cite{iori}, the fundamental hypothesis is here that if one could gather all the information held by all agents, one should be able to come up with a price that equals the fundamental value. In other terms, I suppose that the market is efficient in the usual meaning. However, one agent alone never has all the information, and thus is not able to come up with the fundamental value by itself alone. Only the financial investors taken as a whole are able to fairly price an asset. More precisely, if V is the fundamental value of a firm, I suppose that $V$ is close to the amount estimated by each agent $V_i = V + v_i$. This fundamental hypothesis implies then that the expected value of $v_i$ is zero by aggregation of all agents $i$. From this perspective, market efficiency means here the capacity of the market pricing process to reveal the fundamental value that would be known by financial investors as a whole.

Furthermore, I assume that the market clearing price is fixed according to some rule based upon excess of demand or supply. In particular, when they are equal, the market price does not change. Any supply (demand) excess will lead to a decreasing (increasing) change of the market price, while the change process stops whenever the market reaches an down (up) limit when all the agents wish to sell (buy). In fact, modeling the market microstructure is beyond the scope of this paper and is left to a further exercise \cite{y1,y2}.  

\subsection{Common beliefs and the market pricing process}

The approach consists in modeling agents which get and process information in some iterative learning way. This process can drive the actual stock price to move substantially away from its fundamental value. Under certain circumstances, the market seems then inefficient. I will show why and for what reasons a market that is expected to be efficient in treating all personal information can lead instead to a stock price that reflects collective beliefs that diverge from the fundamental value. 

In the first case the market seems efficient whereas in the second case, it seems inefficient, even though similar mechanisms are in play. In fact, in the first case, the market is a machine that discloses the true value of a firm, whereas in the second case, the market is a mechanism that produces baseless financial representations of that value.

My simplified model aims to describe in a coherent and homogenous way the formation of speculative bubbles and market crashes enlightening  the role of common beliefs in the process of stock prices formation.  The goal of this paper is  to illustrate the conceptual contribution of sociophysics to the formation of speculative bubbles.

\subsection{Modeling the Learning Process}

Consider a population of N agents, each having a different opinion based on its information about a given stock in a given market. I suppose that the aggregation of all the information held by every agent would provide the fundamental stock price. 

In order to describe a complex reality, two essential processes are distinguished in individual decision making. Every agent is supposed to go first into its own personal analysis using its own information around selling or buying the stock. Then, because its information is incomplete, it will adjust its own opinion to take into account those of other agents. But, in order to avoid mimicking potentially bad decisions from others, it will proceed sequentially through successive steps and vary the group of agents that it consults. Each agent keeps reproducing this process until it gets a clear-cut buying or selling signal.

I further suppose that all agents have the same power of influence on each other. There is no leader or guru. In addition,  buying or selling decisions are all made simultaneously by all agents. Accordingly, one agent cannot postpone its decision to wait for a favorable time to buy or sell. The iterative process is imposed to all agents. Such iterative process is then exogenous, not endogenous, although it is internal to the agent community.

The first mechanism starts with the market opening, each agent having made its own personal decision to buy or sell. The opening price results from all these decisions. Then a quest for the fundamental value begins through a sequence of local optimizations in which agents will collectively determine their choice by a rule of majority. This is why the model is iterative and describes an interactive learning process by the decisions of other agents.

At the end of the day, when the market closes, two configurations are possible. Either there is still a split between the agents about their decision to buy or sell, or all agents end  up with the same decision. In the first case, the stock Exchange closes up or down, this change being justified by the fundamental stock value. In the second case, the stock cannot trade because all agents are either buyers or sellers, the stock has then attained its up or down limit. The next morning, every agent adjusts its position to take into account the latest closing stock price. In the following, I shall see how a market can prove efficient or inefficient depending on conditions that are  specified below

\section{Distinctive stock market dynamics}

\subsection{The case of efficient market}

Consider a stock whose fundamental value in a given day is above its market price. For instance, if $60\%$ of information gives a bullish signal and $40\%$  of information gives a bearish signal, there will be a $20\%$  discrepancy around the fundamental value. However, no single agent knows all the information required to value the stock. On the contrary, each agent has only two guesses or positions to be chosen, i;e., bullish or bearish. Hence at the opening time, $60\%$  of agents would buy and $40\%$  would sell.

I introduce then a partition of the trading day in a certain number of interactions. Time is measured by these interactions. Thus, between the beginning ($t=0$) and the end of the day ($t=T$), at each interaction time, there is a proportion of buyers  $p_t$ and sellers $1-p_t$, no inactive agents being included. At the beginning of the day, there are $p_0$  buyers (in the previous example $p_0 = 60\%$) and $1-p_t$ sellers.

The dynamic  to compute $p_t$, i.e., the proportion of buyers at each time $t$ as a function of the initial proportion of buyers  $p_0$ has thus to defined. As explained before, agents sequentially adjust their decision by observing decisions by others. To do that, each agent interact within a certain group of agents around it. The parity of the size of this group is crucial. For sake of simplicity, I consider here only the two cases of  groups of size three and four.

\paragraph{Groups of odd size}

in the case of odd size with three agents, each agent compares its decision with 2 other agents taken randomly among all agents. Agents are thus grouped by three. Inside each of these groups, a common decision is made according to the  local majority of the initial decisions of the three agents. If two agents are buyers (sellers) and one is seller (buyer), the seller (buyer) changes its mind and decide to buy as the others wish to do. The following relation holds then between $p_{t+1}$ and $p_t$,

\begin{equation}
\label{P3}
P_3(p_t)\equiv p_{t+1} = p_t^3 + 3 \, p_t^2 \, (1-p_t) \  .
\end{equation}

Thanks to this relation, the evolution of the share of buyers $p_t$ with respect to the time t can be observed during a trading day. At the end of the day, I suppose that all agents make their own decisions to buy or sell the next morning, taking into account the latest closing stock price combined with their respective private informations.
 
The dynamic produced by Eq. (\ref{P3}) is monitored by its fixed points, which are obtained by solving the  equation $p_{t+1} = p_t$. The solutions are $p_S = 0$,  $p_{c,3} = 0.50$ and $p_B= 1$. The first fixed point results from a drop in the stock price until a state denoted the down limit because all agents conclude with the same decision to sell. The last fixed point results from the rise of the stock price until a symmetric state is reached with an up limit. The fixed point $p_{c,3} = 0.50$ results from a perfect equilibrium of the market with exactly as many buyers as sellers. The stock price remains stable, and so does the fundamental value.

\paragraph{Stability of equilibria}

\begin{figure}[h]
\begin{center}
\includegraphics*[width=9cm,height=6cm]{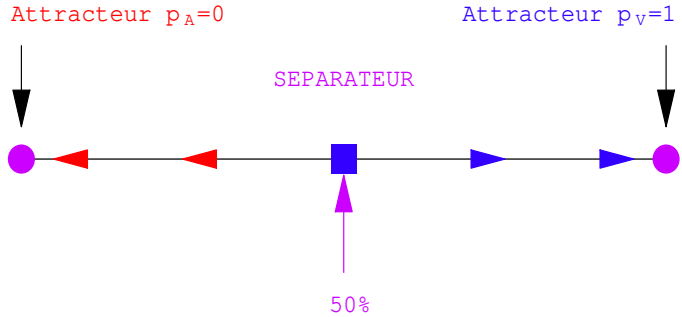}
\caption{An efficient market : 2  attractors at 0 and 1, and a separator at  $0$.}
\label{marche1}
\end{center}
\end{figure}

\begin{figure}[h]
\begin{center}
\includegraphics*[width=9cm,height=9cm]{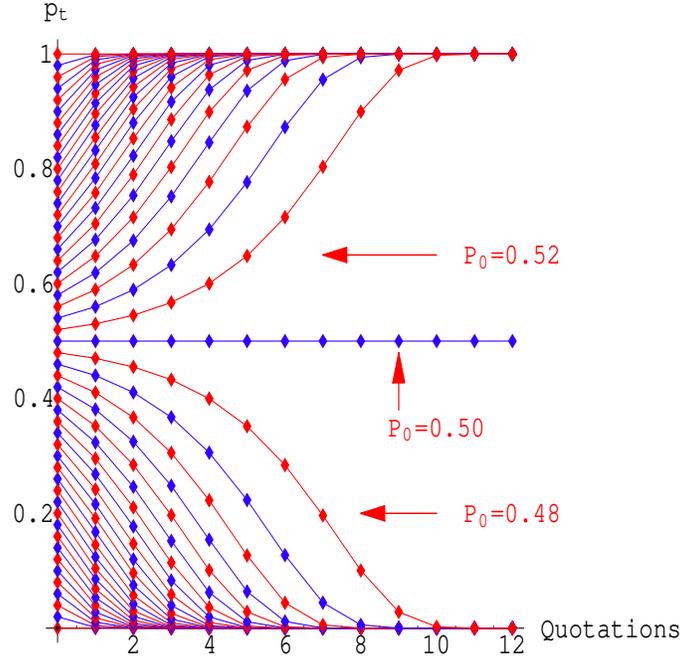}
\caption{An efficient market : variation of $p_t$  for groups of size 3 in function of $p_0$.}
\label{marche2}
\end{center}
\end{figure}

\begin{figure}[h]
\begin{center}
\includegraphics*[width=9cm,height=8cm]{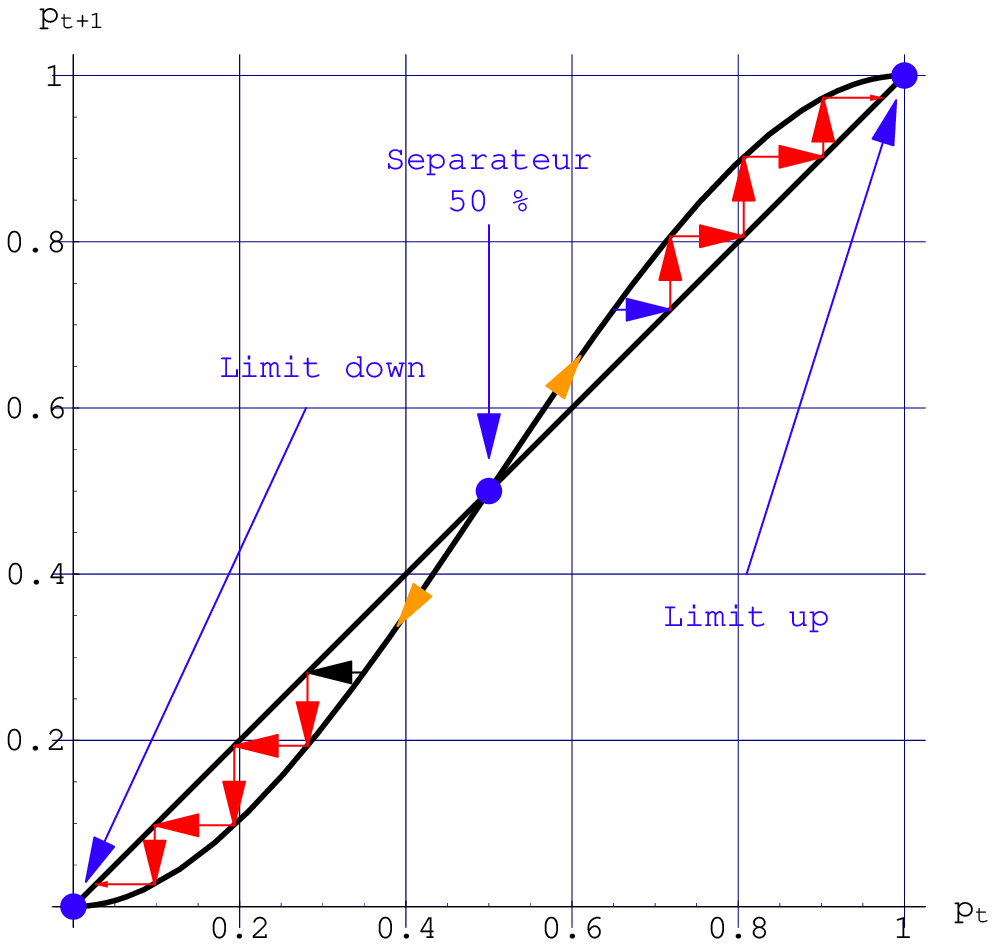}
\caption{An efficient market: variation of $p_{t+1}$ in function of $p_t$  for groups of size 3.}
\label{marche3}
\end{center}
\end{figure}

Studying the stability of  the fixed points $p_{c,3}$ is found to unstable whereas $p_S$ and $p_B$ are stable. If $p_t$ goes slightly away from $p_S$ or $p_B$, Eq. (\ref{P3})  makes it go back towards it. On the contrary, if $p_t$ moves a bit from the fixed point  $p_{c,3}$, Eq. (\ref{P3}) takes it even further in the same direction. Around $p_{c,3}$ a buyers excess will generate the appreciation of the stock price, while a sellers excess will result in a stock price decrease. Figures (\ref{marche1}, \ref{marche2}, \ref{marche3})  illustrate different aspects of this phenomenon. 

Figure (\ref{marche1}) illustrates how the market treats fragmented pieces of information in the right direction. If at the opening time, there are more bullish than bearish signals, the stock will go up during the day. The market is efficient at the informational level. Figure  (\ref{marche2}) shows the variation of $p_t$  during the day for different starting shares of buyers among the whole population  with $p_0 = 0.48. 0.50. 0.52, 0.54,\cdots, 0.98, 1$. For instance, $p_0 = 0.48$ yields 12 interactions to have the market reach its state of down limit.  The proportion $p_0=0$ is then reproduced endlessly. 

Symmetrically, 12 iterations are also required to reach the state of limit up but starting from  $p_0 = 0.52$.
Only at exactly $p_0 = 0.50$ the proportion is invariant under an infinite number of iterations. 

Finally, Figure \ref{marche3} illustrates the variations of $p_{t+1}$ in function of $p_t$. The arrows indicate the direction of the buying flow. For an initial proportion below $p_{c,3} = 0.50$, $p_t$  goes towards the attractor state of down limit, while it goes towards the attractor state of up limit for a proportion greater than $p_{c,3} = 0.50$.

I have therefore obtained a market that works efficiently by transforming information into prices by successive steps, this transformation leading the stock price toward its fundamental value. If the mechanism starts with private information leading to a majority of agents being bullish (bearish), it implies by assumption that the fundamental value of the stock is lower (higher) than its market price. Thus, as interactions go on, the resulting stock market pricing process should and will, push the market higher (lower) in the direction of the fundamental value. However, if an attractor is reached, the dynamic loses contact with the fundamentals of the stock, and the market gets to its up or down limits. It is important to notice that, in this model, a state of down limit down can happen without any previous rise of the stock price.

 What happens the next morning when the market opens again? After a rise (drop) justified by the fundamental value of the stock, a lower proportion of buyers (sellers) will usually occur, leading the stock price to reach the separator that is its fundamental value. This adjustment allows avoiding the perpetuation of up or down limits over time that would lead eventually to a market crash or a bubble.

\paragraph{Alarm signals for market breakdowns}

In this model, the discrepancy between the fundamental value and the market price determines the number of interactions required to reach either one of the two attractors. The more the initial proportion of buyers is far away from the separator, the smaller this number is, and the more likely is a crash or a bubble to occur. This number can be computed to be approximately as

\begin{equation}
\label{nombre}
n \simeq - \frac{ \ln \mid p_{c,3} - p_0 \mid }{ \ln (3/2) }  
\end{equation}

Above number is approximated up to the greater integer. It is noticed that $n$ is always a small number as Figure (\ref{marche2})) shows. For instance, if $p_0 = 0.45$, the formula (\ref{nombre})  gives $n=8$ meaning that starting from an initial proportion of buyers of 45\% , the market reaches a state of limit down after only 8 transactions.  In other terms, if there is a 5\%  lack of buyers of the stock, the driving mechanism of the stock price will push the market lower until it reaches a state of  limit down after 8 interactions. This implies a cumulative amplification of the difference between the number of sellers and the number of buyers. The number of buyers decreases according to Eq.  (\ref{P3}). Starting from the initial value $p_0 = 0.45$ I successively get as proportions of buyers the values $p_1 = 0.42$, $p_2 = 0.39$, $p_3 = 0.34$, $p_4 = 0.26$, $p_5 = 0.17$, $p_6 = 0.08$, $p_7 = 0.02$ and $p_8 = 0.00$. At the 8th interaction, there is no more buyers, the state reaches the down limit and the process stops.

\begin{figure}[h]
\begin{center}
\includegraphics*[width=9cm,height=8cm]{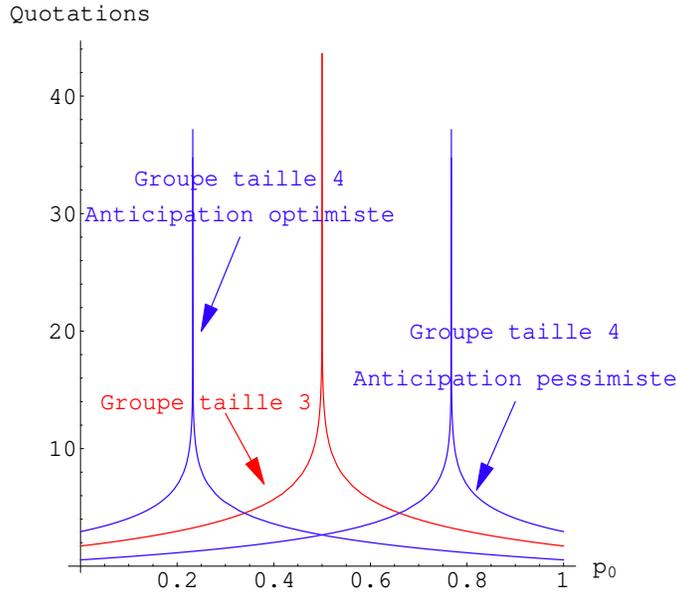}
\caption{Number of iterations required to reach an attractor from an initial proportion of  $p_0$ for groups of size 3 and 4. In the last case both optimistic and pessimistic anticipations are shown.}
\label{marche4}
\end{center}
\end{figure}

The formula (\ref{nombre})  therefore enables to simulate the behavior of buyers and to predict the occurrence of a brutal drop of the market price (up to its down limit). However, the formula is less efficient for proportions $p_0$ of  buyers that would be already closed to attractors, depicting a market process potentially stopped. In such situation that is too close to a market stop the number of interactions which separates the market from a state of down or up limit has to be evaluated by an explicit iteration of Eq. (\ref{P3}). The Figure (\ref{marche4}) is an illustration of this phenomenon. The number of iterations does not become equal to zero for the proportions $p_0 = 0$ and $p_0 = 1$. Actually, in the vicinities of both attractors, the formula (\ref{nombre}) has to be adjusted by the quantity $\ln (1/2) / \ln (3/2) $.

\subsection{The case of market inefficiency}

So far, the model has described an efficient market, since the market mechanism makes the stock price reaching its fundamental value from the confrontation of buying and selling by partially informed agents. Moreover, the market dynamic that results from their aggregation is perfectly symmetric whether the stock price increases or decreases. This shall be no longer the case when groups of even size are considered.

\paragraph{Groups of even size and the formation of a doubt}

LetÕs consider now a different market situation. Instead of interacting with two other agents, every agent  now considers the decisions from three other agents. The market population is then organized by  groups of size 4. In this case, a new possibility emerges when there are as many buyers as there are sellers in a group (2 buyers and 2 sellers). The majority choice cannot solve this situation. Accordingly the agents in such a group must find another way out of this paralyzing doubt. 

I make the assumption that, in a situation of doubt, a collective decision to buy or sell is made based on the collective current consensus for the economical sector the stock belongs to, that is, the socio-psychological mood of the market in that sector. If the consensus is optimistic, the group of agents collectively decide to buy, otherwise it sell. This assumption allows studying the effect of this collective expectation subsuming the current mood of consensus, on the efficiency property of the market. In order to do so, Eq.  (\ref{P3}) needs to be rewritten with a  further determination of  the novel location of the separator.

To take into account the psychological collective mood, I introduce a new parameter $k$ which measures the optimism level for the sector the stock belongs to. Thus, $k=1$ depicts a completely optimistic atmosphere, leading all doubting groups to buy in case of  a local doubt. On the contrary, $k=0$ depicts a completely pessimistic mood, leading people to sell in case of a doubt. More realistically, $k$ can take values between 0 and 1. Following the ongoing changes in the mood, this value can vary day by day, slowly or quickly. The general mood which used to be optimistic may then become abruptly pessimistic or vice-vera. With the introduction of this confidence parameter, Eq.  (\ref{P3}) is generalized as follows,

\begin{equation}
\label{P4}
P_4(p_t)\equiv p_{t+1} = p_t^4 + 4 \, p_t^3 (1-p) + 6 k p_t^2 (1-p_t)^2 \ .
\end{equation}
Like Eq. (\ref{P3}), this equation gives the proportion of buyers $p_{t+1}$ at time $(t+1)$ in function of the proportion of buyers $p_t$ at time $t$.

\paragraph{Changes in the proportion of buyers}

Fom Eq. (\ref{P4}) the equation  $p^* = P_4(p^*)$ yields again the two previous attractors $p_S = 0$  and $p_AB= 1$,  but the separator $p_{c,4}^k$ has shifted from the middle point with $p_{c,4}^k \not= 0.50$) at a value  given by

\begin{equation}
\label{pc4}
p_{c,4}^k = \frac{ (6k-1) - \sqrt{ 13 - 36 k + 36k^2} }{ 6(2k-1) } \ .
\end{equation}
This value depends on $k$, the degree of optimism of the market. The complete optimism ($k=1$) gives,

\begin{equation}
\label{pc4-1}
p_{c,4}^1  = \frac{5 - \sqrt{13}}{6} \simeq 0.23 \ ,
\end{equation}
while the complete pessimism $(k=0)$ gives the separator value,
\begin{equation}
\label{pc4-1}
p_{c,4}^1 = \frac{1 + \sqrt{13}}{6} \simeq 0.77 \ .
\end{equation}

Above two values represent the two bounds between which the separator can vary. This means that, in a frankly optimistic atmosphere, it suffices to have 23\%  of buyers instead of 50\%  in an efficient market, to modify market equilibrium. Reciprocally, a very pessimistic atmosphere requires 77\%  of buyers to influence market equilibrium against  50\%  in an efficient market. As expected, for a balance collective mood, i.e., $k=0.50$ yields $p_{c,4}^{0.50} = 0.50$ as in the previous model. Figure  (\ref{marche5}) shows the  varying range of the separator.

 {\begin{figure}[t]
\begin{center}
\includegraphics*[width=9cm,height=4cm]{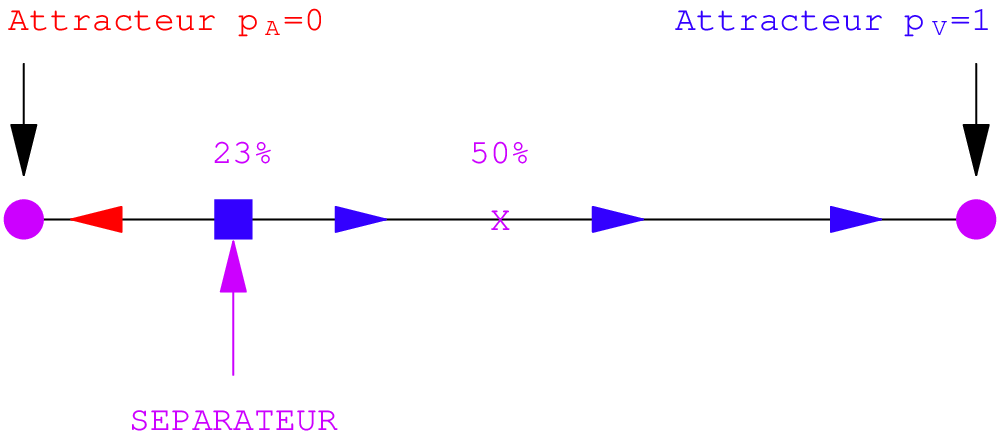}
\includegraphics*[width=9cm,height=4cm]{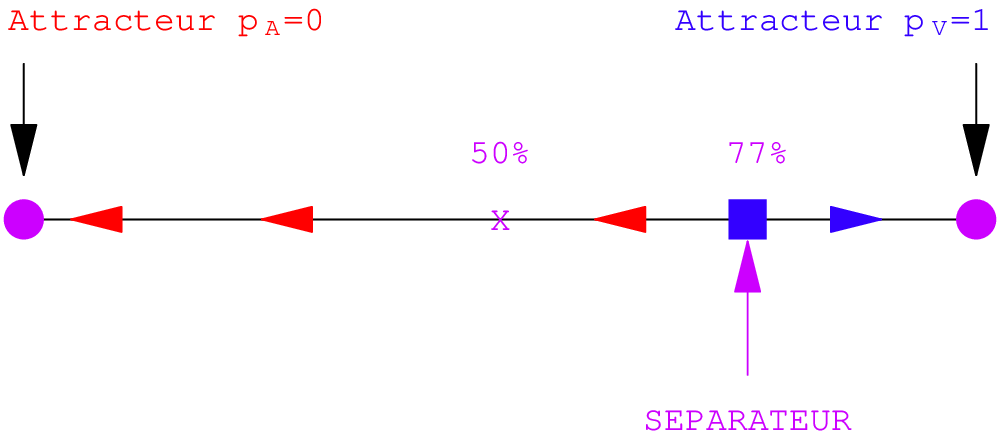}
\caption{Direction of the dynamic of sequential change of the proportion of buyers, with two attractors at 0 and 1 for groups of size 4. On top, the anticipation is fully optimistic with $k =1$, which sets the separator at 23\%. On bottom the anticipation is fully pessimistic with $k = 0$ setting the separator at 77\%.}
\label{marche5}
\end{center}
\end{figure}

What happens to the efficiency property of our market? When $0.50 < k < 1$ the psychological mood is optimistic. Therefore, even if the proportion of buyers $p_t$ is less than 50\%  down to  the limit of 23\%, the power of the collective belief ,which is optimistic with respect to the future of the asset, pushes the stock price up. In the same way, in a negative psychological mood, as long as the proportion of buyers remains below 77\%, the stock price will be driven down.

In other words, a separator away from the middle value 0.50 makes the market inefficient since the stock price can go in an opposite direction to what the majority of agents taken individually would think it should go. Such uncertainty makes the market process dependent on the market atmosphere. The psychological mood may, in certain situations, bias significantly the market dynamic. The aggregation of all pieces of information by the market may then lead to the wrong conclusion by the majority of agents. Figure (\ref{marche7}) illustrates this mechanism.

\begin{figure}[h]
\begin{center}
\includegraphics*[width=9cm,height=8cm]{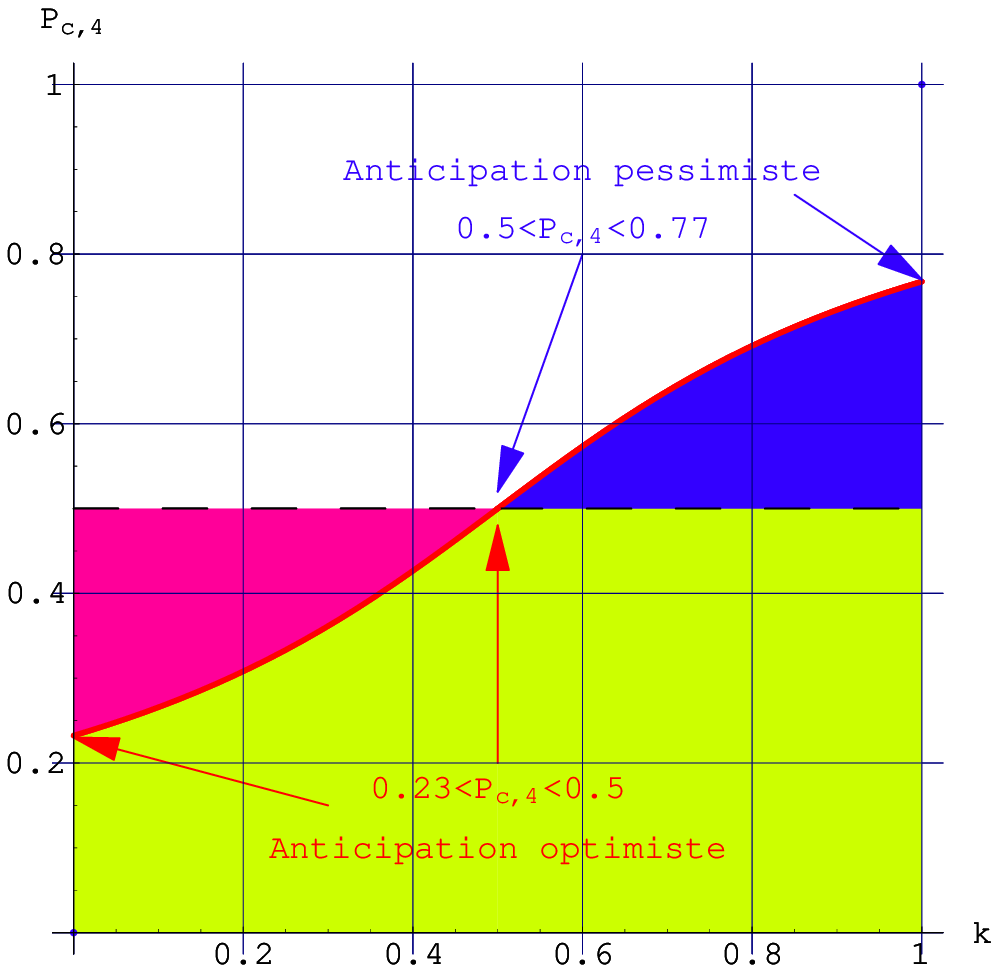}
\caption{Location of the separator $p_{c,4}^k$  as a function of the psychological mood $k$. An optimistic mood ($0.50<k\leq1$)  yields $0.23< p_{c,4}^k < 0.50$ whereas a pessimistic mood ($0 \leq k <0.50$)) gives $0.50 < p_{c,4}^k < 0.77$. The discrepancy between $p_{c,4}^k$ and 0.50 is the source of the market inefficiency as shown in the area on each side of  $p_{c,4}^{0.50} = 0.50$.}
\label{marche6}
\end{center}
\end{figure}   

In particular, the market will push higher an already overvalued stock, as illustrated in Figure (\ref{marche7}), in the case of a totally optimistic common belief ($k=1$). In all situations where the general market mood is determinant, only the neutral value $k = 0.50$ makes the market efficient with $p_{c,4}^{0.50} = 0.50$.

However, the market inefficiency has well defined  limits. For instance, in a speculative bullish market, private information can become so negative that it can reestablish the market efficiency by successfully counterbalancing the bias created by the market atmosphere. In this case, the market will start to be driven down after a continued and unjustified bullish period. The market mechanism eventually overcomes the distortion implied by the psychological anticipation and proves efficient again. This process of readjustment may be partially smooth and stay under control as shown in Figures (\ref{marche7})  and (\ref{marche8}). On these figures, the initial proportion of buyers or sellers are very close to the separator $p_0 \simeq 0.23$ that exhibits the limit of the inefficiency of the market.

\begin{figure}[h]
\begin{center}
\includegraphics*[width=9cm,height=8cm]{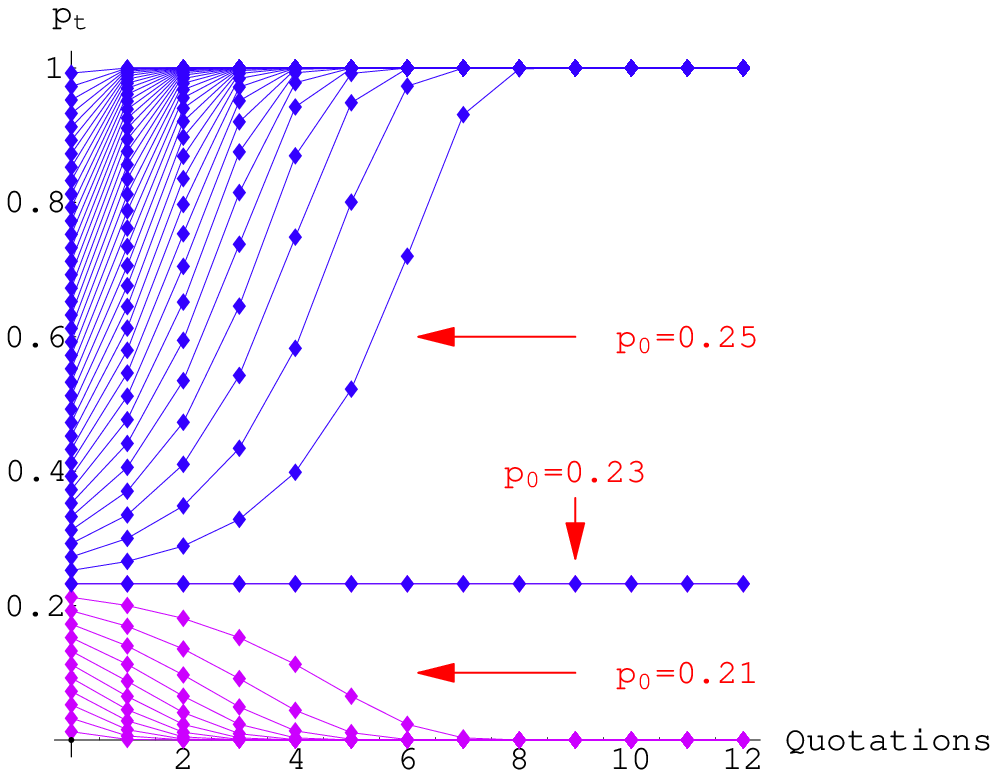}
\caption{Variation of $p_t$  for groups of size 4 in a very optimistic collective mood $(k=1)$ as a  function of the initial proportions of buyers. The initial proportions vary from $p_0 = p_{c,4}^1  + \mu  \times \  \delta  \times \ 0.01$ with $p_{c,4}^1  \simeq 0.23$ and $\delta = 0. 1, 2,\cdots, 77$ for $\mu=+1$ and  $\delta=0. 1, 2, ..., 23$ for $\mu=-1$. }
\label{marche7}
\end{center}
\end{figure}   

\begin{figure}[h]
\begin{center}
\includegraphics*[width=9cm,height=8cm]{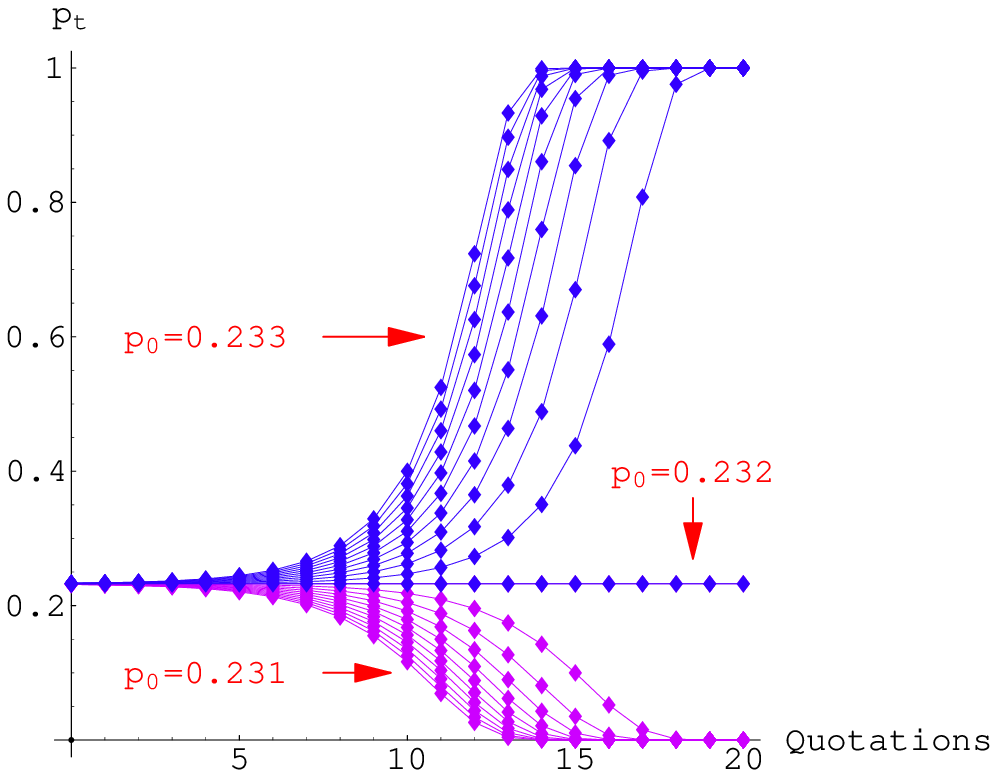}
\caption{Variation of p(t) for groups of size 4 in a very optimistic mood (k=1) as a function of initial proportions of buyers which vary from $p_0=p_{c,4}\pm 0.001 \times \delta$ with $p_{c,4}^1  \simeq 0.23$  and $\delta = 0, 1, 2, É,10$. That is the limit of malfunctioning of the market: a correction in the stock price remains smooth. }
\label{marche8}
\end{center}
\end{figure}

\subsection{Speculative bubbles and crashes}

\paragraph{Change in collective beliefs and sharp corrections}

I have analyzed the limits of the inefficiency of the market in this model. I can now study the conditions for which the market beaks into a crash.  Such a drastic event is naturally obtained within our model. Indeed, I have considered so far the market atmosphere constant with time. However, changes in the general market mood can occur for a variety of reasons with as well its direction as  its intensity.

For instance, in the previous case of a very optimistic mood ($k=1$) with a separator set at $p_{c,4}^1  \simeq 0.23$, I saw that the market can regain its efficiency if it goes too far in the wrong direction. However, it is possible that, before the market starts to correct itself, the general anticipation  brutally changes to shift into a very pessimistic mood ($k=0$) after having been very optimistic. In such a situation, the drop of the stock price is not a smooth process anymore. In that case, the separator jumps at once from $p_{c,4}^1\simeq 0.23$  up to $p_{c,4}^0 \simeq 0.77$  It may then take only two or three update sequences for the market to reach a state of down limit. The extreme speed of the stock move thwarts the rebalancing of the next day and a self-increasing panic appears. Figure (\ref{marche9}) illustrates the evolution of such a scenario for different initial proportions of buyers. 

\begin{figure}[t]
\begin{center}
\includegraphics*[width=9cm,height=8cm]{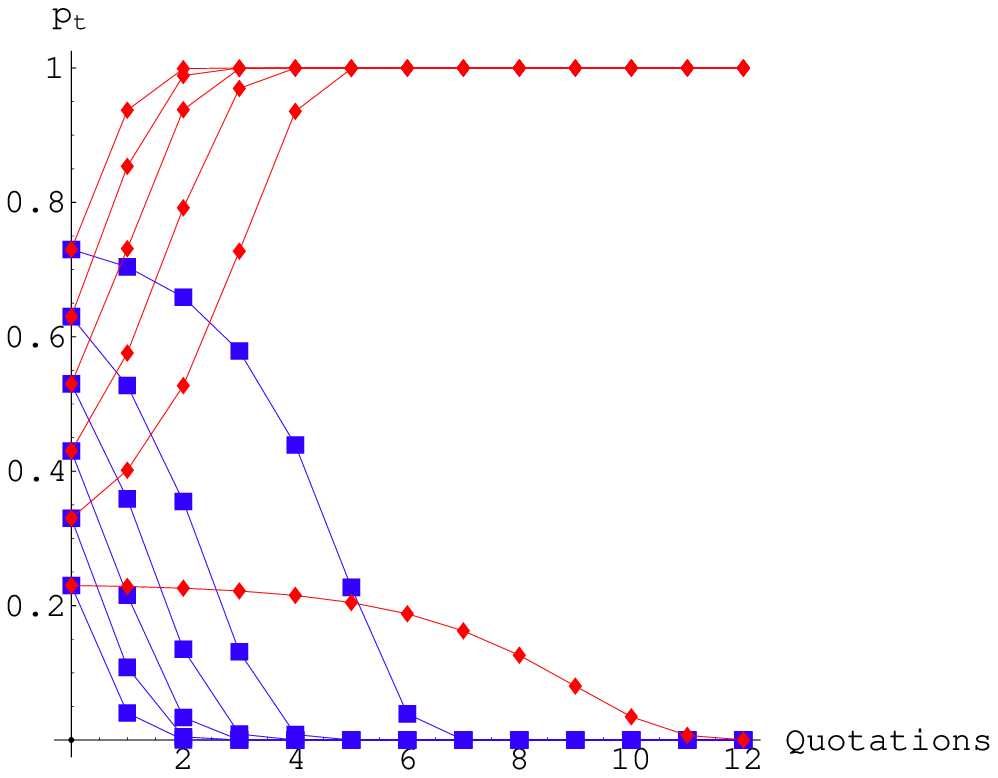}
\caption{Variation of $p_t$ for groups of size 4 with both optimistic and pessimistic moods. Different initial proportions of buyers are shown, which all lead to a bearish market for $k=0$. For $k=1$ on the contrary, they all lead to a bullish market except for $p_0= 0.23$. A change in the psychological anticipation triggers the crash.}
\label{marche9}
\end{center}
\end{figure}

In the case of an optimistic expectations, the market regains its previous bullish dynamics except for $p_0=0.23$, which leads to a smooth decrease of the stock price. On the contrary, if expectations are pessimistic, the decrease of the stock price is brutal, especially when the initial proportion of buyers is small compared to 0.5. To illustrate this mechanism of rupture, consider a market with an initial proportion of buyers $p_0=0.43$ or 0.33 under an optimistic climate. Even if less than half operators are bullish, the stock price will go up. But if the general mood of operators happens to change, the stock can violently crashes as seen from Figure (\ref{marche9}).

\paragraph{Conclusive remarks}

I have shown how a simple sociophysics approach can disclose important features which are instrumental in the making of  the stock market dynamics. In particular, these features shed light on the influence of collective beliefs on stock price formation paving the way to the study of mental representations in play in speculative mechanisms. However, the same features produce the limit of common beliefs on biaising  the market efficiency.

Two distinctive mechanisms characterizing the stock market have been singled out. The first one exhibits the aggregation of the agent  private informations through the market  to treat all scattered and incomplete  informations to reproduce the fundamental value.  At this stage I have been able to build an efficient market. The second mechanism is more subtle trying to embody the effect of  of the collective beliefs regarding the future stock price evolution into the process of the agent decision making. On this basis I have shown  that in case a gap occurs between a strong belief and the fundamental value, the market becomes inefficient with the emergence of a speculative bubble. However, the general belief can only bias the market up to a certain point beyond which private information is so unbalanced that its drive back the stock price toward its fundamental value. This process, if accompanied by a shift in the general belief, can be extremely brutal and leads to a market crash. 

In upcoming work I will introduce different types of agents like contrarians and inflexibles in the market dynamics. They were found to have drastic effect on the landscape of the opinion dynamics of social systems \cite{contra, inflexe,mosco}. I hope my  results could be helpful for a better understanding of  the fundamental trends of the market even though they are not an accurate description of the reality. 
 
\section*{Acknowledgement}

I would like to thank Elie Galam and Yuri Biondi for very helpful comments on my first draft of the manuscript. I want also to thank Wei-Xing Zhou for offering me the opportunity to present this work at the International Conference on Econophysics held in Shanghai in June 2011.


\end{document}